\documentclass[aps,prl,amsmath,amssymb,superscriptaddress,showpacs,balancelastpage,10pt,reprint]{revtex4-1}

\usepackage[dvips]{graphicx}
\graphicspath{{Figures/}}
\newcommand{\setup}[1]{{\textsf{#1}}}

\begin{document}

\title{Field locked to Fock state by quantum feedback with single photon corrections}

\author{X.~Zhou}
\author{I.~Dotsenko}
\email{igor.dotsenko@lkb.ens.fr}
\author{B.~Peaudecerf}
\author{T.~Rybarczyk}
\author{C.~Sayrin}
\author{S.~Gleyzes}
\author{J.M.~Raimond}
\author{M.~Brune}
\affiliation{Laboratoire Kastler-Brossel, ENS, UPMC-Paris 6, CNRS, 24 rue Lhomond, 75005 Paris, France}
\author{S.~Haroche}
\affiliation{Laboratoire Kastler-Brossel, ENS, UPMC-Paris 6, CNRS, 24 rue Lhomond, 75005 Paris, France}
\affiliation{Coll\`ege de France, 11 place Marcellin Berthelot, 75005 Paris, France}

\date{\today}


\begin{abstract}
Fock states with photon numbers $n$ up to 7 are prepared on demand in a microwave superconducting cavity by a quantum feedback procedure which reverses decoherence-induced quantum jumps. Circular Rydberg atoms are used as quantum non-demolition sensors or as single photon emitter/absorber actuators. The quantum nature of these actuators matches the correction of single-photon quantum jumps due to relaxation. The flexibility of this method is suited to the generation of arbitrary sequences of Fock states.
\end{abstract}




\maketitle

The preparation of non-classical field states and their protection against decoherence is an important aspect of quantum physics and of its application to information science. Among the proposed methods, including error correction~\cite{Steane96}, decoherence-free subspaces~\cite{DFS} and reservoir engineering~\cite{Poyatos96}, quantum feedback~\cite{Wiseman94,Doherty00,Wiseman09} is particularly promising. Its principle is to drive the quantum system towards a target state by the repeated action of a sensor-controller-actuator loop. The sensor performs quantum measurements and provides information to the controller. Taking into account the back-action of the measurement, the controller then estimates the system's state and programs the actuator to drive the system as close as possible to the target. A given feedback algorithm can in principle protect a wide class of target states. The operating point can be changed at any time and the system driven through a programmed trajectory in its Hilbert~space.

Photon number (Fock) states are appealing targets for quantum feedback operation. They combine a theoretical and intuitive simplicity ($|n\rangle$ is an eigenstate of the field Hamiltonian with $n$ quanta) with intrinsically non-classical features (their Wigner functions take negative values for $n\ge 1$). When coupled to an environment, Fock states lose rapidly their non-classicality with a time constant $T_n=T_c/n$, where $T_c$ is the lifetime of the field mean energy~\cite{HarocheBook}. Hence, large $n$ Fock states, whose decay time scale is much shorter than $T_c$, are important tools for the exploration of decoherence at the quantum/classical boundary~\cite{Brune08}.

The most intuitive algorithm for quantum feedback generation and protection of Fock states combines single photon emitters/absorbers together with a Quantum Non-Demolition (QND) photon counting sensor. Starting from vacuum, repeated photon emissions make the field climb the ladder of Fock states until the sensor recognizes that the target is reached. When an environment-induced single-photon quantum jump occurs, the sensor detects it and the controller triggers the action of either an emitter or absorber to correct for it.

Resonant actuators and QND sensors have already been used separately to prepare Fock states in a time short compared to $T_n$. Repeated photon emission by resonant actuators has been achieved in cavity-~\cite{Walther00} and circuit-QED experiments~\cite{Martinis08}. QND photon counting by repeated atomic sensor measurements has collapsed a coherent field into a Fock state with random $n$ and revealed its subsequent quantum jumps~\cite{Guerlin07}. Single-photon-actuators and QND sensors have never been combined in a feedback loop. The only steady-state quantum feedback procedure so far has stabilized Fock states with a coherent source as actuator~\cite{Sayrin11}. The performance of this scheme was limited by the inability of the classical actuator to correct for a quantum jump in a single step.

    \begin{figure}
     \includegraphics[width=8cm]{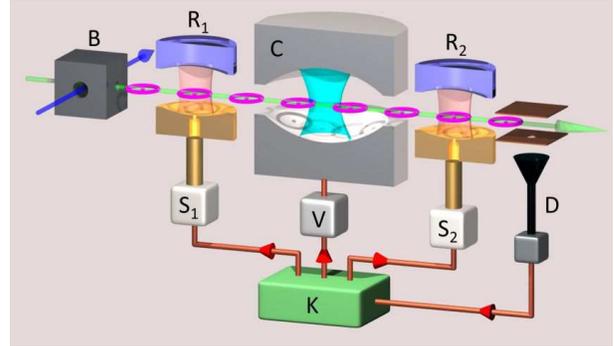}
     \caption{Scheme of the experimental set-up.}
     \label{fig_setup}
    \end{figure}

We report here the implementation of a quantum feedback algorithm for the preparation and protection of Fock states up to $n=7$ in a microwave cavity QED experiment using circular Rydberg atoms as QND sensors and as single-photon actuators~\cite{RMP}. The experimental set-up is depicted in Fig.~\ref{fig_setup}. The microwave field (frequency $\nu =51$~GHz) is stored in the superconducting Fabry-Perot cavity \setup{C} ($T_c=65$~ms) cooled at 0.8~K (average number of blackbody photons $n_{th}=0.05$)~\cite{Kuhr07}. Rydberg atoms are prepared in \setup{B} into the circular state $|g\rangle$ of principal quantum number 50 by a pulsed process repeated at $T_a=82\ \mu$s time intervals. This laser- and r.f.-induced process also selects the atomic velocity ($v= 250$~m/s). The small number of atoms in each sample obeys a Poisson law.

A voltage \setup{V} applied across the cavity mirrors tunes the atoms in or out of resonance with \setup{C} by Stark-shifting the transition between $|g\rangle$ and the nearest higher-lying circular state $|e\rangle$. After crossing \setup{C}, the detection of each atom in $|e\rangle$ or $|g\rangle$ by state-selective field-ionization in \setup{D} provides binary information.

The cavity \setup{C} is placed between two low-$Q$ cavities \setup{R$_1$} and \setup{R$_2$} used to manipulate the atomic state by classical resonant microwave pulses (sources \setup{S$_1$} and \setup{S$_2$}). The experiment is controlled by a computer \setup{K} (ADwin~Pro-II), which processes binary information from \setup{D} and uses it to adjust for each sample the settings of \setup{S$_1$}, \setup{S$_2$} and \setup{V}.  When \setup{K} decides to send a sensor in \setup{C}, \setup{V} is set to make the atom-field interaction dispersive and \setup{S$_1$} and \setup{S$_2$} are set for $\pi/2$ pulses. The \setup{R$_1$}-\setup{R$_2$} combination is then a Ramsey interferometer providing QND information on $n$ by measuring the dispersive phase shift experienced by the atom in \setup{C}~\cite{Guerlin07}. When \setup{K} decides to send an emitting actuator, it applies with \setup{S$_1$} a $\pi$-pulse realizing the $|g\rangle\rightarrow|e\rangle$ transformation in \setup{R$_1$} and it sets \setup{V} so that the atom interacts resonantly with \setup{C} during an adjustable time. The source \setup{S$_2$} is switched off and the atom is directly detected in \setup{D}. For an absorbing actuator, \setup{S$_1$} and \setup{S$_2$} are switched off and the resonant interaction time is controlled by \setup{V}.

The first task of \setup{K} is to estimate the field state after detection of each sample, based on all available information. Since the initial vacuum and the actuators bear no phase information, the field density matrix remains diagonal in the $\{|n\rangle\}$ basis and \setup{K} needs only to update the photon number distribution $p(n)$  after each detection. From $p(n)$, \setup{K} evaluates the distance $d=\sum_n (n-n_t)^2p(n)=\Delta n^2+(\overline n-n_t)^2$ to the target Fock state $|n_t\rangle$ ($\Delta n^2$ and $\overline n$ are the photon number variance and mean value, respectively).

After detection of a sensor in state $j=0$ or 1 ($|e\rangle$ or $|g\rangle$, respectively), $p(n)$ becomes, according to Bayes' law and within a normalization factor, $p(n)\times \pi_s(j|n)$, where $\pi_s(j|n)$ is the conditional probability for detecting the sensor in $j$  when there are $n$ photons. Ideally, $\pi_s(j|n)=[1+\cos(\Phi_0 n+\varphi_r-j\pi)]/2$, where $\Phi_0$ is the phase shift per photon  accumulated by the atomic coherence in \setup{C} and $\varphi_r$ the adjustable phase of the Ramsey interferometer~\cite{Sayrin11}. In the experiment, due to various imperfections, $\pi_s(j|n)$ becomes $[1+jb_{s}+c_s\cos(\Phi_0 n+\varphi_r-j\pi)]/2$, where the offset $b_s=0.02\pm0.002$ and the contrast $c_s=0.75\pm 0.03$ are determined in calibration experiments.

The resonant actuators and \setup{C} perform a Rabi oscillation between the joint states $|e,n\rangle$ and $|g,n+1\rangle$ at a frequency proportional to $\sqrt{n+1}$. After detection of an actuator, prepared in $j$ and detected in $k$ ($j,k=0,1$ for $|e\rangle$ and $|g\rangle$, respectively), the photon number distribution $p(n)$ becomes, within a normalization, $p(n+j-k)\times \pi_a(j,k|n+j-k)$, where $\pi_a(j,k|n+j-k)$ is the conditional probability of the $j\rightarrow k$ transition in the field of $n+j-k$ photons. Ideally, $\pi_a(j,k|n+j-k)=\{1+\cos[ \Omega_0t_j\sqrt{n-k+1}+(j-k)\pi]\}/2$. Here, $\Omega_0/2\pi=47.9$~kHz is the vacuum Rabi oscillation for an atom at cavity center; $t_j=t_e$ or $t_g$ where $t_e$ and $t_g$ are effective interaction times taking into account the atomic motion through the Gaussian structure of the mode in \setup{C}. Experimental imperfections affect the contrast of these oscillations in an $n$-dependent way and slightly offset them. The actual  $\pi_a(j,k|n)$ are obtained by recording Rabi oscillations in calibration experiments.

The state estimation must also take into account the measured detection efficiency ($\eta_d=0.25$). For instance, if no atom is detected, it is possible that the sample was actually empty [$p(n)$ being then unchanged] or that it contained one, or even two undetected atoms. The controller \setup{K} updates $p(n)$ as a weighted average of the modified photon number distributions corresponding to all possible events leading to the actual detection. The probabilities of these events are inferred by Bayes' law, knowing $\eta_d$ and the mean atom number $m$ in a sample \cite{Dotsenko09}. Undetected sensors have no effect on state estimation, since a QND atom does not modify $p(n)$ on the average. Undetected actuators must be considered, the modified $p(n)$ being a weighted average of the ones which would be obtained if the missed atom would have been detected in $|e\rangle$ or $|g\rangle$. If two actuators are crossing \setup{C} together (0, 1 or 2 being detected), the updating formulas take into account the modified Rabi oscillation and the possibility of simultaneous two-photon emission or absorption.

The state estimation includes cavity damping during the time interval $T_a$ towards the thermal equilibrium at the mirrors temperature~\cite{Dotsenko09}. Finally, \setup{K} takes into account the feedback loop delay, i.~e.~the effect of the yet undetected actuator samples that are on their way between \setup{C} and \setup{D}, by tracing over their states.

Getting information on $p(n)$ requires many sensors (each providing one bit), whereas adding or subtracting a photon ideally demands a single atom. Thus, a feedback loop consists of a sequence of $N_s$ sensor samples followed by $N_c < N_s$ control samples whose operating mode is decided by \setup{K}.

After each detection and state estimation, \setup{K} makes or updates a decision about the fate of all control samples located immediately before \setup{R$_1$} or between \setup{R$_1$} and \setup{C}. In the first case, \setup{K} decides upon the sample mode (sensor, emitter or absorber). For actuators between \setup{R$_1$} and \setup{C}, \setup{K} uses additional information acquired after selection of their modes (emitter/absorber) to decide to let them interact resonantly with \setup{C} as previously planned or to cancel their action by keeping them off-resonance. The controller estimates the average photon number distribution resulting from the pending interactions of the control samples with \setup{C} for each of these choices and selects that corresponding to the minimum distance $d$ to the target.

    \begin{figure}
     \includegraphics[width=\columnwidth]{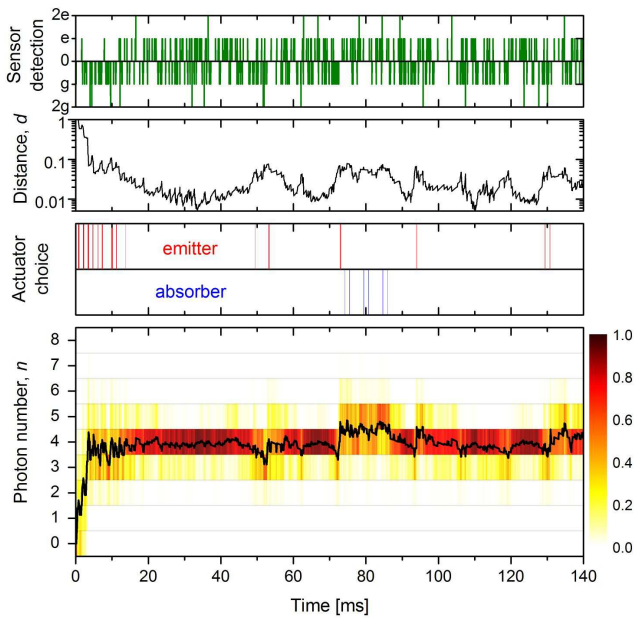}
     \caption{Single realization of the feedback experiment with $n_t=4$. The frames present versus time, from top to bottom, the detected sensor states (upwards bars for $e$, downwards bars for $g$), the distance $d$ to the target, the actuators sent by \setup{K} (red bars for emitters, blue bars for absorbers) and the photon number distribution $p(n)$ inferred by \setup{K} (color/gray scale) together with its average value (solid black line).}
     \label{fig_singletrace}
    \end{figure}

The free experimental parameters are the phase-shift per photon $\Phi_0$, the Ramsey phase $\varphi_r$, the actuator interaction times $t_j$, the number of sensor and control samples $N_s$ and $N_c$ and the corresponding average number of atoms per sample $m_s$ and $m_c$. The phase-shift $\Phi_0=0.252\times\pi$~rad, corresponding to an atom-cavity detuning $\delta/2\pi=244$~kHz, is set close to $\pi/4$ allowing \setup{K} to distinguish among eight different photon numbers~\cite{Guerlin07}. The Ramsey interferometer phase $\varphi_r=\pi/2-\Phi_0n_t$ is set by fine Stark-tuning of the atomic frequency. It corresponds ideally to $\pi_s(j|n_t)=1/2$ and provides the best sensitivity to photon number measurements around $n_t$. We chose $m_s=1.3$ and $m_c=0.5$ (the lower value for the control samples reduces the probability of two-photon emission/absorption). The other parameter values, $t_e=1.6\pi/\Omega_0\sqrt{n_t+1}$, $t_g=2.4\pi/\Omega_0\sqrt{n_t}$, $N_s=12$ and $N_c=4$ are optimized by numerical simulations.

The value of  $t_e$ is close to $2\pi/\Omega_0\sqrt{n_t+1}$. This corresponds to a ``trapping state'' condition~\cite{Walther99}, for which an emitting actuator would ideally leave the target state invariant. Due to the finite contrast of the experimental Rabi oscillations, the emission probability does not cancel at the trapping state condition.  Choosing a slightly lower value for $t_e$ maintains a relatively small unwanted emission probability for $n=n_t$ while optimizing the probability of correcting emissions when $n=n_t-1$. Similar arguments explain qualitatively the value of $t_g$, slightly larger than that corresponding to a trapping state condition.

Figure~\ref{fig_singletrace} shows the data of a single realization of the experiment with $n_t=4$. It presents, as a function of time, the detected sensor states, the estimated distance~$d$, the controller decisions to send emitter or absorber actuator samples, and finally the evolution of the photon number distribution estimated by \setup{K} together with its average value. Starting from vacuum at $t=0$, emitting samples are repeatedly sent until $d$ comes close to zero. The photon number distribution is then peaked on $n=n_t$, with $p(n_t)\approx 0.8-0.9$. Around $t=50$~ms, a downwards quantum jump to $n=3$ triggers the sending of few emitter samples, which rapidly restore the target state. Close to $t=70$~ms, another downwards jump is over-corrected, leading to $n=5$. Absorbers are then sent until restoration of the target. Four thousand similar trajectories have been recorded for each value of $n_t$ from~1 to~7.

    \begin{figure}
     \includegraphics[width=7cm]{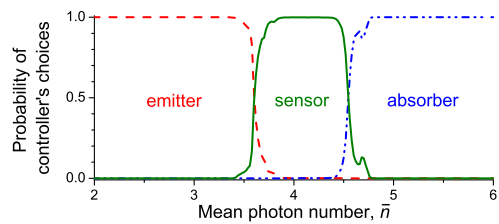}
    \caption{Probabilities of the choices made by \setup{K} for the control atoms (emitter: dashed red line; sensor: solid green line; absorber: dash-dot blue line)  as a function of the estimated mean photon number $\overline{n}$ (data inferred from 4000 realizations of the experiment over 140~ms with $n_t=4$).}
     \label{fig_controllerchoices}
    \end{figure}

In order to get an intuitive insight into the workings of the feedback, we plot for $n_t=4$ in Fig.~\ref{fig_controllerchoices} the fractions of emitters, absorbers and sensors chosen by \setup{K} in the control samples, versus the mean photon number $\overline n$ estimated at a given time. The mere inspection of this figure leads to a simple rule. When $\overline n<n_t-0.4$ (resp. $\overline n>n_t+0.6$), \setup{K} essentially programs emitter (resp. absorber) samples and for $n_t-0.4 \le \overline n \le n_t+0.6$ it rather decides to send sensor samples, which do not affect $\overline n$ on average, but contribute to reducing $\Delta n$ and thus $d$. The domain of $\overline n$ values in which sensors are preferred has a $\delta n\approx 1$ width and is centered on a value slightly larger than $n_t$, reflecting the fact that field relaxation alone reduces $\overline n$. The fractions of actuators and sensors vary rapidly at the boundary of these domains, in narrow ranges of $\overline n$ values. Our feedback algorithm thus essentially operates according to the simple intuitive procedure outlined in the introduction. When it estimates that  $\overline n$ falls outside the $\delta n=1$ range around $n_t$, it concludes that a quantum jump is likely to have happened and attempts to correct for it by adding or subtracting a photon.


    \begin{figure*}
     \includegraphics[width=\textwidth]{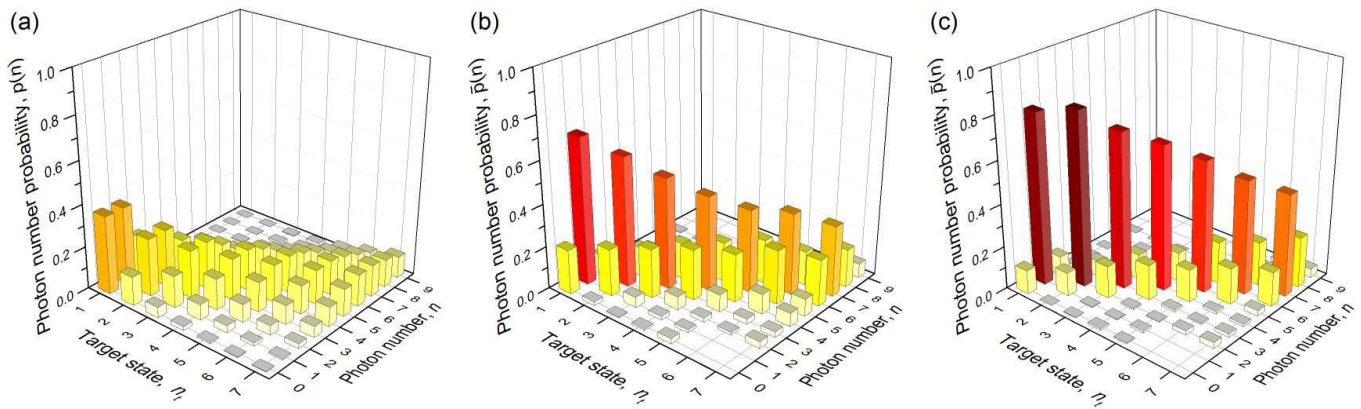}
     \caption{Histograms of photon number distribution as a function of the target photon number $n_t$. (a) Reference Poisson distribution with $n_t$ photons on the average. (b) Photon number distribution $\overline p(n)$ measured by an independent QND process after interrupting the feedback loop at $t=140$~ms. (c) Photon number distribution $\overline p(n)$ measured when \setup{K} estimates that $p(n_t)>0.8$. For (b) and (c), $\overline p(n)$ is measured from $n_0$ to $n_0+7$, with $n_0=0$ for $n_t\le 5$, $n_0=2$ for $n_t>5$.}
     \label{fig_histogramspns}
    \end{figure*}

The performance of the feedback procedure is obtained by reconstructing the average $\overline p(n)$ distribution, independently from the estimation made by \setup{K}, and by comparing it to the reference Poisson distribution with $n_t$ photons on the average shown in Fig.~\ref{fig_histogramspns}(a). The loop is interrupted at $t=140$~ms. We then send a few QND sensor samples with Ramsey phase settings optimizing the discrimination of 8 photon numbers~\cite{Guerlin07}. From 4000 realizations of this experiment, we reconstruct $\overline p(n)$, plotted  in Fig.~\ref{fig_histogramspns}(b) between $n_0$ and $n_0+7$ around $n_t$, by a maximum likelihood procedure~\cite{Brune08}. The measured $\overline p(n_t)$, i.~e.~the fidelities with respect to the target state, are about twice those of the corresponding Poisson distribution and the $\overline p(n)$ distributions are clearly sub-Poissonian.

    \begin{figure}[b]
     \includegraphics[width=\columnwidth]{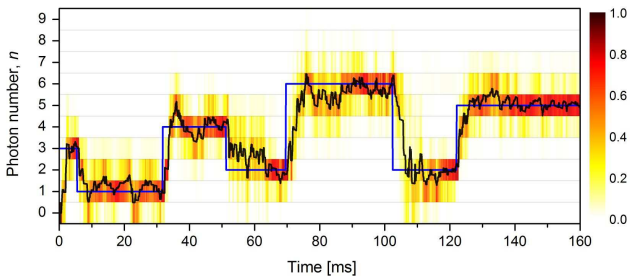}
     \caption{Programmed sequence of Fock states. The target $n_t$ varies stepwise as indicated by the thin line. The $p(n)$ distribution inferred by \setup{K} is shown in color/gray scale, together with its average value (thick line).}
     \label{fig_sequentialpreparation}
    \end{figure}

A better performance is obtained when using the state estimation by \setup{K} for interrupting the feedback at a proper time. The histograms in Fig.~\ref{fig_histogramspns}(c) present for each $n_t$ the $\overline p(n)$ distributions obtained with 4000 sequences interrupted when \setup{K} estimates that $p(n_t)>0.8$. The obtained fidelity is close to the expected 0.8 value up to $n_t=3$ and remains always larger than that of the histograms in Fig.~\ref{fig_histogramspns}(b). From a statistical analysis of these trajectories, we also infer the convergence time towards the target. For $n_t=3$, 63\% of the trajectories have reached the 0.8 fidelity threshold within $27$~ms. This time is twice as short as that observed with classical actuators~\cite{Sayrin11}.

To illustrate the flexibility of the procedure, \setup{K} is programmed to change $n_t$ in real time according to a preset sequence, $\{3,1,4,2,6,2,5\}$ for the single trajectory presented in Fig.~\ref{fig_sequentialpreparation}. The controller switches from one $n_t$ value to the next when it estimates that $p(n_t)> 0.8$. It accordingly adapts the Ramsey phase $\varphi_r$ to the new target. The $p(n)$ distribution and its average $\overline n$ (thick black line) follow rapidly the changes in the target (thin blue line).

The operation of the feedback mechanism can also be viewed as that of a quantum micromaser~\cite{Walther99} pumped by the actuators and actively locked to a Fock state. Based on information provided by the sensors, the controller adjusts in real time the atomic medium gain by choosing the fraction of emitters and absorbers. In ref.~\cite{Sayrin11} the controller was rather injecting coherent microwave pulses into the cavity. The mismatch between the classicality of the source and the quantumness of the target limited the procedure to low photon number states. Here in contrast, the quantum nature of the source allows us to correct more rapidly and precisely decoherence-induced quantum jumps. We lock efficiently the micromaser operation to higher Fock states, making them available for fundamental studies and quantum information experiments.

\begin{acknowledgments}
The authors acknowledge fruitful discussions with P. Rouchon and support from  European Research Council (DECLIC project), the European Community (AQUTE project) and the Agence Nationale de la Recherche (QUSCO-INCA project).
\end{acknowledgments}

\bibliographystyle{plain}

\end{document}